\documentstyle[preprint,pre,aps,epsf]{revtex}
\begin{document}
\draft

\tightenlines

\title{Closed description of stationary flames}

\author{Kirill~A.~Kazakov\thanks{E-mail: $Kirill@theor.phys.msu.su$}}

\address{Department of Theoretical Physics, Physics Faculty,
Moscow State University, \\ 117234, Moscow, Russian Federation}

\maketitle

\begin{abstract}
The problem of non-perturbative description of stationary flames with
arbitrary gas expansion is considered. A general method for deriving
equations for the flame front position is developed. On the basis of the
Thomson circulation theorem an implicit integral of the flow equations is
constructed. With the help of this integral the flow structure near the
flame front is determined, and a closed system of equations for the flame
front position and a set of auxiliary quantities is obtained. This system
is shown to be quasi-local, and its transverse representation is found.
This representation allows reduction of the system to a single equation
for the flame front position. The developed approach is applied to the
case of zero-thickness flames.
\end{abstract}


\unitlength=1pt

\noindent

\section{Introduction}

The process of flame propagation presents an extremely complicated
mathematical problem. The governing equations include the nonlinear flow
equations for the fuel and the products of combustion, as well as
the transport equations governing the heat conduction
and species diffusion inside the flame front. Fortunately, in practice,
an inner flame scale defined by the latter processes is large compared
to the flame front thickness, implying that the flame can be considered
as a gasdynamic discontinuity. The initial problem is reduced thereby to
a purely hydrodynamic problem of determining the propagation of a surface
of discontinuity in an incompressible fluid, the laws of this propagation
being determined by the usual Navier-Stokes and continuity equations
complemented by the jump conditions at the surface, expressing the mass
and momentum conservation across the flame front. The asymptotic methods
developed in Refs.~\cite{siv1,matalon,pelce} allow one to express these
conditions in the form of a power series with respect to the small flame
front thickness.

Despite this considerable progress, however, a closed theoretical
description of the flame propagation is still lacking. What is meant by
the term ``closed description'' here is the description of flame dynamics
as dynamics of the flow variables {\it on} the flame front surface.
Reduction of the system of bulk equations and jump conditions, mentioned
above, to this ``surface dynamics'' implies solving the flow equations for
the fuel and the combustion products, satisfying given boundary conditions
and the jump conditions at the flame front, and has only been carried out
asymptotically for the case $\theta\to 1,$ where $\theta$ is the gas
expansion coefficient defined as the ratio of the fuel density and the
density of burnt matter \cite{siv1,sivclav,kazakov1,kazakov2}.

Difficulties encountered in trying to obtain a closed description of
flames are conditioned by the following two crucial aspects:

(1) Characterized by the flow velocities which are typically well below
the speed of sound, deflagration represents an essentially nonlocal
process, in the sense that the flame-induced gas flows, both up- and
downstream, strongly affect the flame front structure itself. A seeding
role in this reciprocal process is played by the Landau-Darrieus (LD)
instability of zero thickness flames \cite{landau,darrieus}. A very
important factor of non-locality of the flame propagation is the vorticity
production in the flame, which highly complicates the flow structure
downstream. In particular, the local relation between pressure and
velocity fields upstream, expressed by the Bernoulli equation, no longer
holds for the flow variables downstream.

(2) Deflagration is a highly nonlinear process which requires an
adequate non-perturba-\\tive description of flames with arbitrary values
of the flame front slope. As a result of development of the LD-instability,
exponentially growing perturbations with arbitrary wavelengths make any
initially smooth flame front configuration corrugated. Characteristic
size of the resulting "cellular" structure is of the order of the cutoff
wavelength $\lambda_c \sim 20 L_{\rm f}$ given by the linear theory of
the LD-instability \cite{pelce}; $L_{\rm f}$ is the flame front thickness.
The exponential growth of unstable modes is ultimately suppressed by the
nonlinear effects. Since for arbitrary $\theta$ the governing equations do
not contain small parameters, it is clear that the LD-instability can only
be suppressed by the nonlinear effects if the latter are not small, and
therefore so is the flame front slope.

The stabilizing role of the nonlinear effects is best illustrated
in the case of stationary flame propagation. Numerical experiments on 2D
flames with $\theta = 6-8$ show that even in very narrow tubes (tube
width of the order $\lambda_c$), typical values for the flame front slope
are about $1.5-2.0$ \cite{bychkov1}. Nonlinearity can be considered small
only in the case of small gas expansion, $\theta\to 1,$ where one has the
$O(\theta-1)$ estimate for the slope, so that it is possible to derive
an equation for the flame front position in the framework of the perturbation
expansion in powers of $(\theta - 1)$ \cite{sivclav,kazakov1,kazakov2}.

This perturbative method gives results in a reasonable agreement with
the experiment only for flames with $\theta \le 3,$ propagating in very
narrow tubes (tube width of the order $\lambda_c$), so that the front slope
does not exceed unity. Flames of practical importance, however, have
$\theta$ up to 10, and propagate in tubes much wider than $\lambda_c.$
As a result of development of the LD-instability, such flames turn out to
be highly curved, which leads to a noticeable increase of the flame velocity.
In connection with this, a natural question arises whether it is possible
to develop a non-perturbative approach to the flame dynamics, closed in the
sense mentioned above, which would be applicable to flames with
arbitrary gas expansion.

A deeper root of this problem is the following {\it dilemma:} On the one
hand, flame propagation is an essentially non-local process [see the point
(1) above], on the other, this non-locality itself is determined by the
flame front configuration and the structure of gas flows near the front,
so the question is whether an explicit bulk structure of the flow downstream
is necessary in deriving an equation for the flame front position. In other
words, we look for an approach which would provide the closed description
of flames more directly, without the need to solve the flow equations
explicitly.

The purpose of this paper is to develop such approach in the
stationary case.

The paper is organized as follows. The flow equations and related results
needed in our investigation are displayed in Sec.~\ref{flowequations}.
A formal integral of the flow equations is obtained in Sec.~\ref{integral}
on the basis of the Thomson circulation theorem. With the help of this
integral, an exact flow structure near the flame front is determined in
Sec.~\ref{structure}. Using the jump conditions for the velocity
components across the flame front, this result allows one to write down
a system of equations for the flow variables and the flame front position,
which is closed in the above-mentioned sense. This is done in the most
general form in Sec.~\ref{gen}. It is remarkable that this system turns out
to be quasi-local, which implies existence of the transverse representation
derived in Sec.~\ref{transverse}. The developed approach is applied
to the particular case of zero-thickness flames in Sec.~\ref{zero}, where
a single equation for the flame front position is derived. The results
obtained are discussed in Sec.~\ref{discussion}.

\section{Integral representation of flow dynamics}

As was mentioned in the point (1) of Introduction, an important
factor of the flow non-locality downstream is the vorticity production in
curved flames, which highly complicates relations between the flow variables.
In the presence of vorticity, pressure is expressed through the velocity
field by an integral relation, its kernel being the Green function of the
Laplace operator. It should be noted, however, that the jump condition for
the pressure across the flame front only serves as the boundary condition
for determining an appropriate Green function, being useless
in other respects. Thus, it is convenient to exclude pressure from
our consideration from the very beginning. The basis for this
is provided by the well-known Thomson circulation theorem. Thus,
we begin in Sec.~\ref{flowequations} with the standard formulation of
the problem of flame propagation, and then construct a formal implicit
solution of the flow equations with the help of this theorem
in Sec.~\ref{integral}.

\subsection{Flow equations}\label{flowequations}

Let us consider a stationary flame propagating in an initially uniform
premixed ideal fluid in a tube of arbitrary width $b.$ To make the
mathematics of our approach more transparent, we will be dealing below
with the 2D case. Let the Cartesian coordinates $(x,z)$ be chosen so that
$z$-axis is parallel to the tube walls, $z = - \infty$ being in the fresh
fuel. It will be convenient to introduce the following dimensionless
variables
$$(\eta,\xi )=(x/R,z/R)\,, \  (w,u) =({\rm v}_{x}/U_{\rm f}\,,
{\rm v}_{z}/U_{\rm f})\,,$$
$$\Pi =(P-{P}_{\rm f}) /{\rho }_{-}{U_{\rm f}}^{2},$$
where $U_{\rm f}$ is the velocity of a plane flame front, $P_{\rm f}$ is the
initial pressure in the fuel far ahead of the flame front, and $R$
is some characteristic length of the problem (e.g., the cut-off
wavelength). The fluid density will be normalized on the fuel
density $\rho_{-}.$ As always, we assume that the process of flame
propagation is nearly isobaric. Then the velocity and pressure
fields obey the following equations in the bulk
\begin{eqnarray}\label{flow1}
\frac{\partial v_i}{\partial\zeta_i} &=& 0\,,
\\ v_i\frac{\partial v_k}{\partial\zeta_i}
&=& - \frac{1}{\rho}\frac{\partial\Pi}{\partial\zeta_k}\,,
\quad k = 1,2,
\label{flow2}
\end{eqnarray}
\noindent
where $ (\zeta_1,\zeta_2) = (\eta,\xi),\, (v_1,v_2) = (w,u),$ and
summation over repeated indices is implied.

Acting on Eq.~(\ref{flow2}) by the operator
$\varepsilon_{kl}\partial/\partial\zeta_l,$ where
$\varepsilon_{ik} = - \varepsilon_{ki},\  \varepsilon_{12} = + 1,$
and using Eq.~(\ref{flow1}), one obtains a 2D version of the Thomson
circulation theorem
\begin{eqnarray}\label{thomson}
v_i\frac{\partial\sigma}{\partial\zeta_i} = 0\,,
\end{eqnarray}
\noindent
where
$$\sigma \equiv \frac{\partial u}{\partial\eta}
- \frac{\partial w}{\partial\xi}\,.$$
According to Eq.~(\ref{thomson}), the voticity $\sigma$ is conserved
along the stream lines. As a simple consequence of this theorem,
one can find the general solution of the flow equations upstream.
Namely, since the flow is potential at $\xi = - \infty$
($u = V = {\rm const},$ $w = 0,$ where $V$ is the velocity of the flame
in the rest frame of reference of the fuel), it is potential for every
$\xi<f(\eta),$ $f(\eta)$ denoting the flame front position. Therefore,
\begin{eqnarray}\label{solup1}
u &=& \sum\limits_{n= - \infty}^{+ \infty}
~u^{(n)} \exp\left\{\frac{\pi}{b}(|n|\xi + i n\eta)\right\}\,,
\\ \label{solup2}
w &=& \hat{H}(u - u^{(0)})\,,
\end{eqnarray}
\noindent
where the linear Hilbert operator $\hat{H}$ is defined by
\begin{eqnarray}\label{hilbert}
\hat{H}\exp(i k\eta) = i\chi(k)\exp(i k\eta)\,,
\quad k\ne 0\,,
\end{eqnarray}
and $$\chi(x) =
\left\{
\begin{array}{cc}
+1,& x>0\,,\\
-1,&  x<0\,.
\end{array}
\right.
$$
The boundary condition $w = 0$ at the tube walls ($\eta = 0, b$)
implies that the coefficients $u^{(n)}$ are all real,
\begin{eqnarray}\label{ureal}
u^{(n)*} = u^{(n)}\,.
\end{eqnarray}
\noindent
It will be shown in the next section how the Thomson theorem can be
used to obtain a formal integral of the flow equations downstream.

\subsection{Integration of the flow equations}\label{integral}

Consider the quantity $$a_i(\zeta) = \triangle^{-1} v_i
= \int\limits_{\Sigma}\frac{\ln r}{2\pi} v_i d s\,,$$
where $r$ is the distance from an infinitesimal fluid element $d s$
to the point of observation $\zeta,$
$r^2 = (\zeta_i - \tilde{\zeta_i})^2,$ and
integration is carried over
$\Sigma = \{\tilde{\eta},\tilde{\xi}: \tilde{\xi} > f(\tilde{\eta})\}.$
Taking into account Eq.~(\ref{flow1}), one has for the divergence
of $a_i:$
\begin{eqnarray}\label{diva}
\frac{\partial a_i}{\partial\zeta_i}
= \int\limits_{\Sigma}\partial_i \left(\frac{\ln r}{2\pi}\right) v_i ds
= - \int\limits_{\Sigma}\tilde{\partial}_i
\left(\frac{\ln r}{2\pi}\right)  v_i ds
= - \int\limits_{\Sigma}\tilde{\partial}_i
\left(\frac{\ln r}{2\pi} v_i\right) ds
= - \int\limits_{\Lambda}\frac{\ln r}{2\pi} v_i d l_i\,,
\end{eqnarray}
\noindent
where $\tilde{\partial_i} \equiv \partial/\partial\tilde{\zeta}_i,$
$\Lambda$ boundary of $\Sigma,$ and $d l_i $ its element.

Next, let us calculate
$\varepsilon_{ik}\partial_k \varepsilon_{lm}\partial_l a_m.$
Using Eq.~(\ref{diva}), we find
\begin{eqnarray}
\varepsilon_{1k}\partial_k \varepsilon_{lm}\partial_l a_m
&=& \frac{\partial}{\partial\xi}\left(\frac{\partial a_2}{\partial\eta}
- \frac{\partial a_1}{\partial\xi}\right)
= \frac{\partial}{\partial\eta}\left(- \frac{\partial a_1}{\partial\eta}
- \int\limits_{\Lambda}\frac{\ln r}{2\pi} v_i d l_i\right)
- \frac{\partial^2 a_1}{\partial\xi^2}
\nonumber\\
&=& - \triangle a_1
- \frac{\partial}{\partial\eta}
\int\limits_{\Lambda}\frac{\ln r}{2\pi} v_i d l_i\,.
\nonumber
\end{eqnarray}
\noindent
Analogously,
\begin{eqnarray}
\varepsilon_{2k}\partial_k \varepsilon_{lm}\partial_l a_m
= - \triangle a_2
- \frac{\partial}{\partial\xi}\int\limits_{\Lambda}
\frac{\ln r}{2\pi} v_i d l_i\,.
\nonumber
\end{eqnarray}
\noindent
Together, these two equations can be written as
\begin{eqnarray}
\varepsilon_{ik}\partial_k \varepsilon_{lm}\partial_l a_m
= - \triangle a_i - \partial_i\int\limits_{\Lambda}
\frac{\ln r}{2\pi} v_k d l_k\,.
\nonumber
\end{eqnarray}
\noindent
Substituting the definition of $a_i$ into the latter equation, and
integrating by parts gives
\begin{eqnarray}\label{vint}
v_i &=& - \varepsilon_{ik}\partial_k \varepsilon_{lm}
\partial_l \triangle^{-1} v_m
- \partial_i\int\limits_{\Lambda}\frac{\ln r}{2\pi} v_k d l_k
\nonumber\\
&=& \varepsilon_{ik}\partial_k \int\limits_{\Lambda}\frac{\ln r}{2\pi}
\varepsilon_{lm} v_m d l_l
- \partial_i\int\limits_{\Lambda}\frac{\ln r}{2\pi} v_k d l_k
- \varepsilon_{ik}\partial_k \int\limits_{\Sigma}\frac{\ln r}{2\pi}
\sigma d s\,.
\end{eqnarray}
\noindent
The first two terms on the right of Eq.~(\ref{vint}) represent the
potential component of the fluid velocity, while the third corresponds
to the vortex component. The aim of the subsequent calculation is
to transform the latter to an integral over the flame front surface.
To this end, we will decompose $\Sigma$ into elementary $d s$ as follows.

\begin{figure}
\hspace{3cm} \epsfxsize=10cm\epsfbox{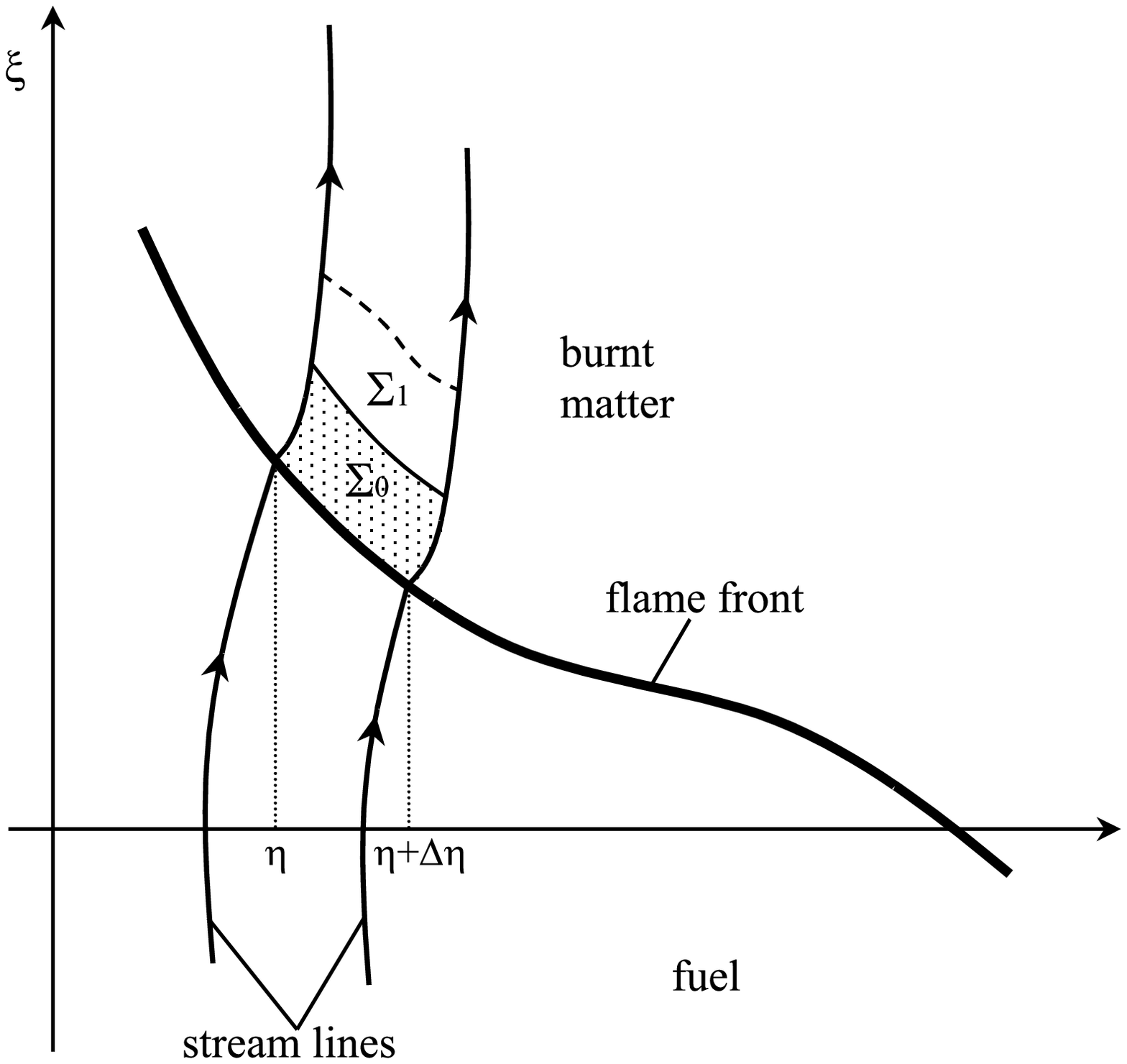}
\caption{Elementary decomposition of the flow downstream.}\label{fig1}
\end{figure}

Let us take a couple of stream lines crossing the flame front at points
$(\eta, f(\eta))$ and $(\eta + \Delta\eta, f(\eta + \Delta\eta))$
(see Fig.~\ref{fig1}).
Consider the gas elements moving between these lines, which cross the
front between the time instants  $t = 0$ and $t = \Delta t.$ During this
time interval, these elements fill a space element $\Sigma_0$ adjacent to
the flame front. For sufficiently small
$\Delta\eta, \Delta t,$ the volume of $\Sigma_0$
$$\Delta s \approx \left|
\begin{array}{cc}
\Delta\eta & f'\Delta\eta \\
w_+ \Delta t & u_+\Delta t \\
\end{array}
\right|
= (u_+ - f'w_+)\Delta\eta \Delta t = v^n_+ N \Delta\eta \Delta t\,,$$
where $$f'\equiv \frac{d f}{d\eta}\,, \quad N\equiv \sqrt{1 + f'^2}\ ;$$
the subscript``$+$'' means that the corresponding quantity is
evaluated just behind the flame front, {\it i.e.,} for $\xi = f(\eta) + 0,$
and $v^n_+ = v_{i+} n_i$ is the normal velocity of the burnt gas,
$n_i$ being the unit vector normal to the flame front
(pointing to the burnt matter).
After another time interval of the same duration $\Delta t,$ the elements
move to a space element $\Sigma_1$ adjacent to $\Sigma_0.$ Since the
flow is incompressible, $\Sigma_1$ is of the same volume as $\Sigma_0.$
Continuing this, the space between the two stream lines turns out to be
divided into an infinite sequence of $\Sigma$'s of the same volume,
adjacent to each other. Thus, summing over all $\Delta\eta,$ the third
term in Eq.~(\ref{vint}) can be written as
\begin{eqnarray}\label{vint1}
- \frac{\varepsilon_{ik}}{2}\partial_k \int\limits_{F}
d l~v^n_+\sigma_{+} K(\eta,\xi,\tilde{\eta})\,,
\end{eqnarray}
\noindent
where $F$ denotes the flame front surface (the front line in our 2D case),
\begin{eqnarray}\label{kernel}
K(\eta,\xi,\tilde{\eta}) &=& \frac{1}{\pi}\lim\limits_{\Delta t\to 0}
\sum\limits_{n=0}^{\infty}
\ln\left\{(\eta - {\rm H}(\tilde{\eta},n \Delta t))^2
+ (\xi - \Xi(\tilde{\eta},n\Delta t))^2\right\}^{1/2}\Delta t
\nonumber\\
&=& \frac{1}{\pi}\int\limits_{0}^{+ \infty}d t
\ln\left\{(\eta - {\rm H}(\tilde{\eta},t))^2
+ (\xi - \Xi(\tilde{\eta},t))^2\right\}^{1/2},
\end{eqnarray}
\noindent
and $({\rm H}(\tilde{\eta},t),\, \Xi(\tilde{\eta},t))$ trajectory of
a particle crossing the point $(\tilde{\eta}, f(\tilde{\eta}))$
at $t = 0.$

Substituting expression (\ref{vint1}) into Eq.~(\ref{vint}) gives
\begin{eqnarray}\label{vint2}
v_i &=& \varepsilon_{ik}\partial_k \int\limits_{\Lambda}\frac{\ln r}{2\pi}
\varepsilon_{lm} v_m d l_l
- \partial_i\int\limits_{\Lambda}\frac{\ln r}{2\pi} v_k d l_k
- \frac{\varepsilon_{ik}}{2}\partial_k \int\limits_{F}
d l~v^n_+\sigma_{+} K(\eta,\xi,\tilde{\eta})\,.
\end{eqnarray}
\noindent
This representation of the flow velocity downstream will be used
in the next section to determine the structure of the vortex
mode near the flame front.

\section{Structure of the vortex mode}\label{structure}

To determine the flame front dynamics, it is sufficient to know
the flow structure near the flame front. As to the flow upstream,
it is described by Eqs.~(\ref{solup1}), (\ref{solup2}) for all
$\xi < f(\eta).$ Given the solution upstream, velocity components
of the burnt gas at the flame front can be found from the jump
conditions which express the mass and momentum conservation across
the front. On the other hand, these components are required to
be the boundary values (for $\xi = f(\eta) + 0$) of the velocity
field satisfying the flow equations. As was shown in the preceding
section, the latter can be represented in the integral form,
Eq.~(\ref{vint2}). Any velocity field can be arbitrarily decomposed
into a potential and vortex modes. The former has the form analogous
to Eqs.~(\ref{solup1}), (\ref{solup2}), namely
\begin{eqnarray}\label{soldown1}
u^p &=& \sum\limits_{n= - \infty}^{+ \infty}
~\bar{u}^{(n)} \exp\left\{ \frac{\pi}{b}(-|n|\xi + i n\eta)\right\}
\,,\\
\label{soldown2}
w^p &=& - \hat{H}(u^p - \bar{u}^{(0)})\,.
\end{eqnarray}
\noindent
Our strategy thus will be to use the integral representation to determine
the near-the-front structure of the vortex mode described by the last
term in Eq.~(\ref{vint2}).

Equation (\ref{vint2}) reveals the following important fact.
Up to a potential, the value of the vortex mode at a given point
$(\eta,\xi)$ of the flow downstream is determined by only one point
in the range of integration over $(\tilde{\eta},t),$ namely that satisfying
\begin{eqnarray}\label{p}
{\rm H}(\tilde{\eta},t) = \eta\,,\quad \Xi(\tilde{\eta},t) = \xi\,.
\end{eqnarray}
\noindent
This is, of course, a simple consequence of the Thomson theorem underlying
the above derivation of Eq.~(\ref{vint2}). It can be verified directly by
calculating the rotor of the right hand side of Eq.~(\ref{vint2}):
contracting this equation with $\varepsilon_{im}\partial_m,$
using $\varepsilon_{ik} \varepsilon_{im} = \delta_{km},$ and taking
into account relation $$\triangle\ln r =
2\pi\delta^{(2)}(\zeta-\tilde{\zeta})\,,$$
one finds
\begin{eqnarray}\label{twod}
\frac{1}{2}\varepsilon_{im}\partial_m\varepsilon_{ik}\partial_k
\int\limits_{F} d l~v^n_+\sigma_{+}(\tilde{\eta})K(\eta,\xi,\tilde{\eta})
= \int\limits_{0}^{+\infty}d t
\int\limits_{F} d l~v^n_+\sigma_{+}(\tilde{\eta})
\delta(\eta - {\rm H}(\tilde{\eta},t))\delta(\xi - \Xi(\tilde{\eta},t))\,.
\end{eqnarray}
\noindent
Since $\delta (x) = 0$ for $x\ne 0,$ the product of $\delta$-functions
picks the point (\ref{p}) out of the whole range of integration in the
right hand side of Eq.~(\ref{twod}).

Now, let us take the observation point $(\xi,\eta)$ sufficiently close
to the flame front, {\it i.e.,} $\xi\approx f(\eta),$ [$\xi>f(\eta)$].
In view of what has just been said, the vortex component for such
points is determined by a contribution coming from the integration over
$\tilde{\eta},t$ near the flame front, which corresponds to small values
of $t.$ Integration over all other $\tilde{\eta},t$ gives rise to a
potential contribution.

The small $t$ contribution to the integral kernel $K(\eta,\xi,\tilde{\eta})$
can be calculated exactly. For such $t$'s, one can write
\begin{eqnarray}\label{stream}
{\rm H}(\tilde{\eta},t)\approx \tilde{\eta} + w_+(\tilde{\eta})t\,,\quad
\Xi(\tilde{\eta},t)\approx f(\tilde{\eta}) + u_+(\tilde{\eta})t\,.
\end{eqnarray}
Let the equality of two fields $\varphi_1(\eta,\xi),\, \varphi_2(\eta,\xi)$
up to a potential field be denoted as $\varphi_1\doteq \varphi_2.$ Then,
substituting Eq.~(\ref{stream}) into Eq.~(\ref{kernel}), and integrating
gives
\begin{eqnarray}\label{kernel1}
K(\eta,\xi,\tilde{\eta}) &=&
\frac{1}{\pi}\int\limits_{0}^{+ \infty}d t
\ln\left\{(\eta - {\rm H}(\tilde{\eta},t))^2
+ (\xi - \Xi(\tilde{\eta},t))^2\right\}^{1/2}
\nonumber\\
&\doteq& \frac{1}{\pi}\int\limits_{0}^{t_0}d t \ln\left\{
v_+^2 t^2 - 2 {\bf r v_+}t + r^2\right\}^{1/2}
\nonumber\\
&=& \frac{1}{\pi}\int
\limits_{-\frac{{\bf r v_+}}{v_+}}^{v_+ t_0 - \frac{{\bf r v_+}}{v_+}}d y
\ln\left\{y^2 - \frac{({\bf r v_+})^2}{v_+^2} + r^2\right\}^{1/2}
\nonumber\\
&=& \frac{1}{\pi v_+}\left[\sqrt{r^2 - \frac{({\bf r v_+})^2}{v_+^2}}
\arctan\frac{y}{\sqrt{r^2 - \frac{({\bf r v_+})^2}{v_+^2}}}
\right.
\nonumber\\&&
\left.
- y + y\ln\left\{y^2 - \frac{({\bf r v_+})^2}{v_+^2} + r^2\right\}^{1/2}
\right]_{-\frac{{\bf r v_+}}{v_+}}^{v_+ t_0 - \frac{{\bf r v_+}}{v_+}}\,.
\end{eqnarray}
\noindent
Here  $v_+$ denotes the absolute value of the velocity field at the flame
front, and $t_0$ is assumed small enough to justify the approximate
equations (\ref{stream}).

As we know, the only point in the range of integration over
$\tilde{\eta},t$ that contributes to the vortex mode is the one
satisfying Eq.~(\ref{p}) or, after integrating over $t,$
\begin{eqnarray}\label{point}
[\eta - \tilde{\eta}] u_+(\tilde{\eta})
- [\xi - f(\tilde{\eta})] w_+(\tilde{\eta}) = 0\,.
\end{eqnarray}
The distance $r$ between this point and the point of observation tends
to zero as the latter approaches to the flame front surface. Thus, taking
$[\xi - f(\eta)]$ small enough, one can make the ratio $t_0/r$ as large
as desired; therefore, the right hand side of Eq.~(\ref{kernel1}) is $\doteq$
\begin{eqnarray}\label{kernel2}
\frac{r}{\pi v_+}
\left\{\sqrt{1 - \left(\frac{{\bf r v_+}}{r v_+}\right)^2}
\left[\frac{\pi}{2} + \arcsin\left(\frac{{\bf r v_+}}{r v_+}\right)
\right] - \frac{{\bf r v_+}}{r v_+} + \frac{{\bf r v_+}}{r v_+}\ln r
\right\} + {\rm TIC}\,,
\end{eqnarray}
\noindent
where ``TIC'' stands for ``Terms Independent of the Coordinates''
$(\eta,\xi).$ Denoting $$\Omega = \frac{{\bf r v_+}}{r v_+}\,,$$ we
finally obtain the following expression for the integral kernel
\begin{eqnarray}
K(\eta,\xi,\tilde{\eta}) \doteq \frac{r}{\pi v_+}
\left\{\sqrt{1 - \Omega^2}
\left(\frac{\pi}{2} + \arcsin\Omega\right)
+ \Omega\ln \frac{r}{e}\right\} + {\rm TIC}\,.
\nonumber
\end{eqnarray}

In order to find the vortex mode of the velocity according to
Eq.~(\ref{vint2}) we need to calculate derivatives of $K.$
Using relation
\begin{eqnarray}\label{aux1}
\frac{\partial\Omega}{\partial\zeta_i}
= \frac{1}{r}\left(\frac{v_{i+}}{v_+} - \Omega\frac{r_i}{r}\right)\,,
\qquad r_i = \zeta_i - \tilde{\zeta}_i\,,
\end{eqnarray}
one easily obtains
\begin{eqnarray}\label{kderiv}
\frac{\partial K}{\partial\zeta_i}
\doteq \frac{1}{\pi v_+}\left(\frac{r_i}{r}
- \Omega\frac{v_{i+}}{v_+}\right)
\frac{\pi/2 + \arcsin\Omega}{\sqrt{1-\Omega^2}}
+ \frac{v_{i+}}{v_+}\ln r\,.
\end{eqnarray}
\noindent
Equation (\ref{kderiv}) can be highly simplified.
Consider the quantity
\begin{eqnarray}
\Upsilon = \frac{\partial}{\partial\zeta_i}\left\{\left(\frac{r_i}{r}
- \Omega\frac{v_{i+}}{v_+}\right)
\frac{\arcsin\Omega - \pi/2}{\sqrt{1-\Omega^2}}
+ \frac{v_{i+}}{v_+}\ln r\right\}\,.
\end{eqnarray}
\noindent
First, we calculate
\begin{eqnarray}\label{aux2}
\frac{\partial}{\partial\zeta_i}
\left(\frac{r_i}{r} - \Omega\frac{v_{i+}}{v_+}\right) =
\left(\frac{\partial_i r_i}{r} - r_i\frac{r_i}{r^3}\right)
- \frac{\partial\Omega}{\partial r_i}\frac{v_{i+}}{v_+}
= \frac{\Omega^2}{r}\,,
\end{eqnarray}
\begin{eqnarray}\label{aux3}
\frac{\partial}{\partial\zeta_i}
\left(\frac{v_{i+}}{v_+}\ln r\right) =
\frac{v_{i+}}{v_+}\frac{r_i}{r^2} = \frac{\Omega}{r}\,.
\end{eqnarray}
\noindent
Second, we note that the vector
$$\beta_i = \left(\frac{r_i}{r} - \Omega\frac{v_{i+}}{v_+}\right)
\frac{1}{\sqrt{1-\Omega^2}}$$
satisfies
$$\beta_i\beta_i = 1, \qquad \beta_i v_{i+} = 0\,,$$
{\it i.e.,} $\beta_i$ is the unit vector orthogonal to ${\bf v}_+\,.$
In addition to that, $\beta_i$ changes its sign at the point defined
by Eq.~(\ref{point}). Therefore, the derivative of $\beta_i,$ entering
$\Upsilon,$ contains a term with the Dirac $\delta$-function. However,
this term is multiplied by $(\arcsin\Omega - \pi/2),$ which turns into
zero together with the argument of the $\delta$-function. Thus, using
Eqs.~(\ref{aux1}),(\ref{aux2}),(\ref{aux3}), one finds
\begin{eqnarray}
\Upsilon = \frac{\Omega^2}{r}\frac{\arcsin\Omega - \pi/2}
{\sqrt{1 - \Omega^2}} + \left(\frac{r_i}{r}
- \Omega\frac{v_{i+}}{v_+}\right)\left[\frac{1}{1 - \Omega^2}
+ \Omega\frac{\arcsin\Omega - \pi/2}{(1 - \Omega^2)^{3/2}}\right]
\frac{\partial\Omega}{\partial\zeta_i}
+ \frac{\Omega}{r} = 0\,.
\nonumber
\end{eqnarray}
\noindent
We conclude that the term
$$\left(\frac{r_i}{r} - \Omega\frac{v_{i+}}{v_+}\right)
\frac{\arcsin\Omega - \pi/2}{\sqrt{1-\Omega^2}}
+ \frac{v_{i+}}{v_+}\ln r $$ in the integral kernel corresponds
to a pure potential. Therefore, we can rewrite Eq.~(\ref{kderiv}) as
\begin{eqnarray}\label{kderiv1}
\frac{\partial K}{\partial\zeta_i}
\doteq \frac{1}{v_+}\left(\frac{r_i}{r} - \Omega\frac{v_{i+}}{v_+}\right)
\frac{1}{\sqrt{1-\Omega^2}} = \frac{\beta_i}{v_+}\,.
\end{eqnarray}
\noindent
Finally, substituting this result into Eq.~(\ref{vint2}), noting that
the vector $\varepsilon_{ki}\beta_k$ is the unit vector parallel to
$v_{i+}$ if $\varepsilon_{ik}r_i v_{k+}>0,$ and antiparallel in the
opposite case, we obtain the following expression for the vortex component,
$v^{v}_i,$ of the gas velocity downstream near the flame front
\begin{eqnarray}\label{vintf}
v^{v}_i = \int\limits_{F} d l~\chi(\varepsilon_{pq}r_p v_{q+})
\frac{v^n_+\sigma_{+}v_{i+}}{2 v^2_+}\,.
\end{eqnarray}
\noindent
Having written the exact equality in Eq.~(\ref{vintf}) we take this
equation as the {\it definition} of the vortex mode.
As a useful check, it is verified in the appendix that the obtained
expression for $v^v_i$ satisfies $$\left(\partial u^v/\partial\eta -
\partial w^v/\partial\xi\right)_+ \equiv \sigma_+\,.$$

\section{Closed description of stationary flames}\label{closed}

After we have determined the near-the-front structure of the vortex
component of the gas velocity downstream, we can write down a closed
system of equations governing the stationary flame propagation.
As was explained in the Introduction, the term ``closed'' means that
these equations relate only quantities defined on the flame front surface,
without any reference to the flow dynamics in the bulk. This system
consists of the jump conditions for the velocity components at the front,
and the so-called evolution equation that gives the local normal fuel
velocity at the front as a function of the front curvature. These
equations (except for the evolution equation) are consequences of the
mass and momentum conservation across the flame front. In Sec.~\ref{gen},
we obtain the closed system in the most general form, without specifying
the form of the jump conditions, and then apply it to the case of zero
thickness flames in Sec.~\ref{zero}.

\subsection{General formulation}\label{gen}

First of all, we need to find the ``on-shell'' expression for the
vortex component, {\it i.e.,} its limiting form for $\xi\to f(\eta) + 0.$
To this end we note that in this limit,
$\chi(\varepsilon_{ik}r_i v_{k+})\to \chi(\eta - \tilde{\eta}),$
therefore, Eq.~(\ref{vintf}) gives
\begin{eqnarray}\label{vinton}
v^{v}_{i+} = \int\limits_{F} d l~\chi(\eta - \tilde{\eta})
\frac{v^n_+\sigma_{+}v_{i+}}{2 v^2_+}\,.
\end{eqnarray}
\noindent
The total velocity field downstream is the sum of the potential mode
having the form Eqs.~(\ref{soldown1}),(\ref{soldown2}), and the vortex
mode. Let us denote the jump $[v_{i+}(\eta) - v_{i-}(\eta)]$ of the gas
velocity $v_i$ across the flame front as $[v_i].$ Here $v_{\pm i}(\eta)
= v_{i}(\eta,f(\eta)\pm 0).$ Then we can write

\begin{eqnarray}\label{general}
v_{i-} + [v_i] = v^p_{i+} +
\int\limits_{F} d l~\chi(\eta - \tilde{\eta})\sigma_{+}
\frac{(v^n_- + [v^n]) (v_{i-} + [v_i])}{2 (v_-^2 + [v^2])}\,.
\end{eqnarray}
\noindent
The jumps $[v_i]$ (as well as $\sigma_+$) are quasi-local functions of the
fuel velocity at the flame front, and of the flame front shape.
Two equations (\ref{general}), together with Eqs.~(\ref{solup2}),
(\ref{soldown2}), and the evolution equation, $v^n_- = v^n_-(f),$
form a closed system of five equations for the five functions
$v_{i-}(\eta), v^p_{i+}(\eta),$ and $f(\eta).$

It should be emphasized that this formulation implies that the potentials
$v_i$ and $v^p_i$ are explicitly expressed in the form of the Fourier
series, Eqs.~(\ref{solup1}) and (\ref{soldown1}), respectively.
Indeed, relations (\ref{solup2}), (\ref{soldown2}) between the flow
variables hold {\it before} the on-shell transition  $\xi \to f(\eta)$
is performed, while Eq.~(\ref{general}) is formulated in terms of
the on-shell variables $v_{\pm i}.$ Thus, the above system is in
fact a system of equations for the front position $f(\eta)$ and
two infinite sets of the Fourier coefficients $u^{(n)}, \bar{u}^{(n)}.$

However, the form of the integral kernel in Eq.~(\ref{general})
makes it possible to avoid this considerable complication, and
to derive a much simpler formulation. This will be demonstrated
in the next section.

\subsection{Transverse representation}\label{transverse}

The aforesaid simplification is based on the fact that Eq.~(\ref{general})
is quasi-local. To show this, we simply increase its differential order
by one. Namely, differentiating this equation with respect to $\eta,$
taking into account relation
\begin{eqnarray}\label{deltad}
\frac{d\chi(x)}{d x}
= 2\delta(x)\,,
\end{eqnarray}
\noindent and performing the trivial integration over $\tilde{\eta},$ we get
\begin{eqnarray}\label{generald}
v'_{i-} + [v_i]' = (v^{p}_{i+})' +
N\sigma_{+}\frac{(v^n_- + [v^n]) (v_{i-} + [v_i])}{v_-^2 + [v^2]}\,.
\end{eqnarray}
\noindent
Given a point at the flame front, Eq.~(\ref{generald}) relates components
of the fuel velocity at this point and their derivatives at the same point.
This property of {\it quasi-locality implies existence of the transverse
representation} of Eq.~(\ref{generald}). We say that a system of equations is
in the transverse representation if all operations involved in this system
(differentiation, integration) are only performed with the first argument
($\eta$) of the flow variables. In other words, the $\xi$-dependence of
the flow variables in such system is purely parametric.

Now, let us show how the system of Eqs.~(\ref{solup2}), (\ref{soldown2}),
and (\ref{generald}) can be brought to the transverse form. All we have to
do is to express the full $\eta$-derivatives in terms of the partial ones.
This is easily done using the continuity equation (\ref{flow1}) and
the potentiality of the fields $v_{i-}, v^p_i$ as follows:
\begin{eqnarray}\label{transr1}
\frac{d u_-}{d\eta} &=&
\left(\frac{\partial u}{\partial\eta}\right)_-
+ f'\left(\frac{\partial u}{\partial\xi}\right)_-
= \left(\frac{\partial u}{\partial\eta}\right)_-
- f'\left(\frac{\partial w}{\partial\eta}\right)_-
\\  \label{transr2}
\frac{d w_-}{d\eta} &=&
\left(\frac{\partial w}{\partial\eta}\right)_-
+ f'\left(\frac{\partial w}{\partial\xi}\right)_-
= \left(\frac{\partial w}{\partial\eta}\right)_-
+ f'\left(\frac{\partial u}{\partial\eta}\right)_-
\\  \label{transr3}
\frac{d u^p_+}{d\eta} &=&
\left(\frac{\partial u^p}{\partial\eta}\right)_+
+ f'\left(\frac{\partial u^p}{\partial\xi}\right)_+
= \left(\frac{\partial u^p}{\partial\eta}\right)_+
- f'\left(\frac{\partial w^p}{\partial\eta}\right)_+
\\  \label{transr4}
\frac{d w^p_+}{d\eta} &=&
\left(\frac{\partial w^p}{\partial\eta}\right)_+
+ f'\left(\frac{\partial w^p}{\partial\xi}\right)_+
= \left(\frac{\partial w^p}{\partial\eta}\right)_+
+ f'\left(\frac{\partial u^p}{\partial\eta}\right)_+
\end{eqnarray}
\noindent
As to Eqs.~(\ref{solup2}), (\ref{soldown2}), they are already transverse.
Finally, the evolution equation has general form
$$v^n_- = 1 + F(u_-,w_-,f'),$$ where $F$ is a quasi-local function
of its arguments, proportional to the flame front thickness, and therefore
can also be rendered transverse.

Thus, the complete system of governing equations can be written as follows:
$$\left(
\begin{array}{rcl}
\label{generaldr}\displaystyle
\frac{\partial u}{\partial\eta}
- f'\frac{\partial w}{\partial\eta} + [u]^{\prime t}
&=& \displaystyle
\frac{\partial u^p}{\partial\eta}
- f'\frac{\partial w^p}{\partial\eta}
+ N\sigma^t\frac{(1 + F^t + [v^n]^t) (u + [u]^t)}{v^2 + [v^2]^t}\,,
\nonumber\\
\displaystyle
\frac{\partial w}{\partial\eta}
+ f'\frac{\partial u}{\partial\eta} + [w]^{\prime t}
&=& \displaystyle
\frac{\partial w^p}{\partial\eta}
+ f'\frac{\partial u^p}{\partial\eta}
+ N\sigma^t\frac{(1 + F^t + [v^n]^t) (w + [w]^t)}{v^2 + [v^2]^t}\,,
\nonumber\\
w &=& \hat{H}(u - u^{(0)})\,,
\nonumber\\
w^p &=& - \hat{H}(u^p - \bar{u}^{(0)})\,,
\nonumber\\
u - f' w &=& N + N F^t\,.
\end{array}
\right)_{\xi = f(\eta)}\eqno (*)$$
\noindent
The superscript ``$t$'' in these equations means that the corresponding
quantity is to be expressed in the transverse form using
Eqs.~(\ref{transr1}) - (\ref{transr4}), yet without setting
$\xi = f(\eta)$ in their arguments. The latter operation is displayed
out the large brackets in ($*$).

The meaning of the transformations performed is that the $\xi$-dependence
of the flow variables is now irrelevant for the purpose of derivation
of equation for the flame front position. Indeed, since all operations
in the set of equations ($*$) are carried through in terms of $\eta$ only,
one can solve these equations with respect to $f(\eta)$ {\it under the large
brackets} signifying the on-shell transition. Furthermore, since the
function $f(\eta)$ is itself $\xi$-independent, we can consider all
the flow variables involved in ($*$) $\xi$-independent (since the resulting
equation for $f(\eta)$ does not contain these variables anyway), and
to omit the large brackets.\footnote{This reasoning has been repeatedly
used in Refs.~\cite{kazakov1,kazakov2,kazakov3} in deriving perturbative
equations for the flame front position in the stationary as well
as non-stationary cases.}

\subsection{Zero-thickness flames}\label{zero}

As an illustration, the developed approach will be applied in this
section to the case of zero-thickness flames.

In this case, the jump conditions for the velocity components have the form
\begin{eqnarray}\label{jump1}
[u] &=& \frac{\theta - 1}{N}\ ,\\
\label{jump2}
[w] &=& - f'\frac{\theta - 1}{N}\,,
\end{eqnarray}
\noindent
while the evolution equation
\begin{eqnarray}\label{evolution}
v^n_- = 1\,.
\end{eqnarray}
\noindent
We see that the jumps are velocity-independent, and $F\equiv 0.$
Also, it follows from these equations that
$$ [v^n] = \theta - 1\,, \quad [v^2] = \theta^2 - 1\,.$$
Next, substituting the Fourier decomposition, Eq.~(\ref{solup1}),
into evolution equation written as
\begin{eqnarray}\label{evolution1}
u_- - f' w_- = N\,,
\end{eqnarray}
\noindent
and taking into account Eq.~(\ref{ureal}), we get
\begin{eqnarray}
u^{(0)} + 2\left\{\sum\limits_{n = 1}^{+ \infty}
~\frac{u^{(n)}b}{\pi n}
\sin\left(\frac{\pi n\eta}{b}\right)
\exp\left(\frac{\pi |n| f(\eta)}{b}\right)\right\}'
= N\,.
\nonumber
\end{eqnarray}
\noindent
Integrating this equation over interval $(0,b)$ gives
$$b u^{(0)} = \int\limits_{0}^{b} d\eta N = b V\,, \qquad
{\rm or}\qquad u^{(0)} = V\,.$$

It remains only to calculate the value of the vorticity at the flame front,
as a function of the fuel velocity. This can be done directly using
the flow equations (\ref{flow1}),(\ref{flow2}), see Ref.~\cite{matalon}.
With the help of Eqs.~(5.32) and (6.15) of Ref.~\cite{matalon}, the
jump of the vorticity across the flame front can be written,
in the 2D stationary case, as
\begin{eqnarray}\label{vort1}&&
[\sigma] = - \frac{\theta - 1}{\theta N}
\left(\hat{D}w_{-} + f'\hat{D}u_{-} + \frac{1}{N}\hat{D}f'\right),
\end{eqnarray}
\noindent
where
\begin{eqnarray}\label{operator}
\hat{D}\equiv \left(w_{-} + \frac{f'}{N}\right)\frac{d}{d\eta}\,.
\end{eqnarray}
Differentiating the evolution equation (\ref{evolution1}), and
writing Eq.~(\ref{vort1}) longhand, expression in the  brackets can be
considerably simplified
\begin{eqnarray}\label{vort2}&&
\hat{D}w_{-} + f'\hat{D}u_{-} + \frac{1}{N}\hat{D}f'\equiv
w_{-}'w_{-} + \frac{(f'w_{-})'}{N} + \frac{(f')^2 u_{-}'}{N} +
f'u_{-}'w_{-} + \frac{N'}{N} \nonumber\\&&
 = \frac{(w_{-}^2)'}{2} + \frac{(u_{-} - N)'}{N}
+ \frac{(N^2 - 1) u_{-}'}{N} + u_{-}'(u_{-} - N) + \frac{N'}{N} =
u_{-}'u_{-} + w_{-}'w_{-}\,.
\end{eqnarray}
\noindent Since the flow is potential upstream, we obtain the
following expression for the vorticity just behind the flame front
\begin{eqnarray}\label{vort3}&&
\sigma_{+} = - \frac{\theta - 1}{\theta N}(u_{-}'u_{-} + w_{-}'w_{-})\,.
\end{eqnarray}

Using Eqs.~(\ref{transr1}), (\ref{transr2}) in Eq.~(\ref{vort3}),
substituting the result into ($*$), and omitting the large brackets
we arrive at the following fundamental system of equations
\begin{eqnarray}\label{main1}&&
\upsilon' - f'\omega' + (\theta - 1)\left(\frac{1}{N}\right)'
= \upsilon^{p\prime} - f'\hat{\Phi}\upsilon^{p}
\nonumber\\&&
- (\theta - 1)\left\{\upsilon\left(
\upsilon' - f'\omega' \right) + \omega \left(\omega' + f'\upsilon'
\right)\right\}\frac{\upsilon
+ (\theta - 1)/N}{\upsilon^2 + \omega^2 + \theta^2 - 1}\,,
\\&&\label{main2}
\omega' + f'\upsilon' - (\theta - 1)\left(\frac{f'}{N}\right)'
= \hat{\Phi}\upsilon^p + f'\upsilon^{p\prime}
\nonumber\\&&
- (\theta - 1)\left\{\upsilon\left(
\upsilon' - f'\omega' \right) + \omega \left(\omega' + f'\upsilon'
\right)\right\}
\frac{\omega
- f'(\theta - 1)/N}{\upsilon^2
+ \omega^2 + \theta^2 - 1}\,,
\\&&\label{main3}
\omega = \hat{H}(\upsilon - V)\,,
\\&&\label{main4}
\upsilon - f' \omega = N \,,
\end{eqnarray}
\noindent
where $\upsilon, \omega, \upsilon^p, \omega^p$ are the $\xi$-independent
counterparts of the flow variables $u, w, u^p, w^p,$ respectively,
and $\hat{\Phi} = - \hat{H}d/d\eta$ is the Landau-Darrieus operator.

It is not difficult to reduce the above system of equations to a single
equation for the function $f(\eta).$ For this purpose it is convenient
to denote $\tilde{\upsilon} = \upsilon - V\,.$ Then Eqs.~(\ref{main3}),
(\ref{main4}) can be solved with respect to $\tilde{\upsilon}$ to give
\begin{eqnarray}\label{usol}&&
\tilde{\upsilon} = (1 - f'\hat{H})^{-1}(N - V) = (1 + f'\hat{H} +
f'\hat{H} f'\hat{H} + \cdot\cdot\cdot)(N - V)\,.
\end{eqnarray}
\noindent
The two remaining Eqs.~(\ref{main1}) and (\ref{main2}) are linear
with respect to $\upsilon^{p}, \hat{\Phi}\upsilon^{p}\,.$ To exclude
$\upsilon^{p}\,,$ we will use a complex representation of these equations.
Namely, we multiply Eq.~(\ref{main2}) by $i$ and add it to Eq.~(\ref{main1}):
\begin{eqnarray}\label{main5}&&
(\upsilon' + i\omega') (1 + i f')
+ (\theta - 1)\left(\frac{1 - i f'}{N}\right)'
= (\upsilon^{p\prime} + i \hat{\Phi}\upsilon^{p}) (1 + i f')
\nonumber\\&&
-  (\theta - 1)\left\{\upsilon\upsilon' + \omega\omega'
+ f'(\omega\upsilon' - \upsilon\omega')\right\}
\frac{\upsilon + i\omega + (\theta - 1)(1 - i f')/N}
{\upsilon^2 + \omega^2 + \theta^2 - 1}\ .
\end{eqnarray}
\noindent
It follows from the definition of the Hilbert operator, Eq.~(\ref{hilbert}),
that $$\hat{H}^2 = - 1\,.$$ Therefore,
$$(1 + i\hat{H})(\upsilon^{p\prime} + i \hat{\Phi}\upsilon^{p}) =
(1 + i\hat{H})(1 - i\hat{H})\upsilon^{p\prime}
= (1 - i^2\hat{H}^2)\upsilon^{p\prime} = 0\,.$$ Thus, dividing
Eq.~(\ref{main5}) by $(1 + i f'),$ acting by the operator
$(1 + i\hat{H})$ from the left, and taking into account that
$$(1 + i\hat{H})(\upsilon^{\prime} + i\omega^{\prime}) =
(1 + i\hat{H})(1 + i\hat{H})\upsilon^{\prime}
= (1 + 2 i\hat{H} + i^2\hat{H}^2)\upsilon^{\prime}
= 2(1 + i\hat{H})\upsilon^{\prime}\ ,$$
we obtain an equation for the flame front position
\begin{eqnarray}\label{main}&&
(1 + i\hat{H})\left\{
2\tilde{\upsilon}' + \frac{\theta - 1}{1 + i f'}
\left(\frac{1 - i f'}{N}\right)'
+  \frac{\theta - 1}{1 + i f'}
\left[(V + \tilde{\upsilon})\tilde{\upsilon}'
+ (\hat{H}\tilde{\upsilon})(\hat{H}\tilde{\upsilon}')
\right.
\right.
\nonumber\\&&
\left.
\left.
+ f'(\tilde{\upsilon}'\hat{H}\tilde{\upsilon}
- (V + \tilde{\upsilon})\hat{H}\tilde{\upsilon}')\right]
\frac{V + (1 + i\hat{H})\tilde{\upsilon} + (\theta - 1)(1 - i f')/N}
{(V + \tilde{\upsilon})^2 + (\hat{H}\tilde{\upsilon})^2 + \theta^2 - 1}
\right\} = 0\,,
\end{eqnarray}
\noindent
where $\tilde{\upsilon}$ is given by Eq.~(\ref{usol}).

Equation (\ref{main}) provides the closed description of the stationary
zero-thickness flames in the most convenient form. If desired, one can
bring it to the explicitly real form by extracting the real or imaginary
part.\footnote{Both ways, of course, lead to equivalent equations: Acting
by the operator $\hat{H}$ on the real part gives the imaginary one, and
{\it vice versa.}} Account of the effects inside the thin flame front
changes the right hand side of this equation to $O(\varepsilon),$
$\varepsilon = L_{\rm f}/R$ being the relative flame front thickness.

\section{Discussion and conclusions}\label{discussion}

The results of Sec.~\ref{closed} solve the problem of closed description
of stationary flames. The set of equations $(*)$ obtained in
Sec.~\ref{transverse} gives a general recipe for deriving
equations governing the ``surface dynamics'' of the fields $u, w, u^p, w^p,$
or, more precisely, their $\xi$-independent counterparts, $\upsilon,
\omega, \upsilon^p, \omega^p,$ and the function $f(\eta)$ -- the flame
front position. Following this way, we derived an equation for the front
position of zero-thickness flames, Eq.~(\ref{main}). This
equation is universal in that any surface of discontinuity, propagating
in an ideal incompressible fluid, is described by Eq.~(\ref{main})
whatever internal structure of this ``discontinuity'' be.
The latter shows itself in the $O(\varepsilon)$-corrections to this equation,
where $\varepsilon$ is the relative thickness of discontinuity. In the
case of premixed flame propagation, these corrections can be found
using the results of Refs.~\cite{matalon,pelce} following the general
prescriptions of Sec.~\ref{transverse}. It is clear that independently
of the specific form of $\varepsilon$-corrections, the set $(*)$ ultimately
reduces to a single equation for the function $f(\eta),$ since equations
of this set are linear with respect to the field $\upsilon^p\,.$
The latter can be eliminated, therefore, in the same way as in
Sec.~\ref{zero}.

It is interesting to trace the influence of the boundary conditions
on the form of Eq.~(\ref{main}). By itself, expression (\ref{vintf})
for the vortex mode is completely ``covariant'', {\it i.e.,}
it has one and the same vector form whatever boundary conditions. The jump
conditions Eqs.~(\ref{jump1}), (\ref{jump2}) also can be rewritten in an
explicitly covariant form as the conditions on the normal and tangential
components of the velocity. It is the structure of the potential mode
upstream and downstream, given by Eqs.~(\ref{solup1}), (\ref{soldown1}),
respectively, which is directly affected by the boundary conditions for
the velocity field. Thus, it is the boundary conditions which dictate
the way the jump conditions appear in the first two equations of the
set ($*$), as well as the form of the third and forth equations.

Finally, it is worth to compare the results obtained in this paper
and in Refs.~\cite{kazakov1,kazakov2} where an equation describing
stationary flames with arbitrary gas expansion was derived under assumption
that there exists a local relation between the
pressure field and a potential mode of the velocity downstream,
expressed by the Bernoulli-type equation.\footnote{This assumption goes back
to the work \cite{zhdanov} where it was introduced in investigating the
nonlinear development of the LD-instability.} This assumption was proved
in Refs.~\cite{kazakov1,kazakov2} in the framework of perturbative expansion
with respect to $(\theta - 1)$ up to the sixth order.
Comparison of Eq.~(\ref{main}) and Eq.~(40) of Ref.~\cite{kazakov2} now
shows that this assumption is generally wrong (that the two equations are
not equivalent can be easily verified, for instance, considering the large
flame velocity limit investigated in detail in Sec.~V of
Ref.~\cite{kazakov2}). As we saw in Sec.~\ref{integral}, the use of the
Thomson theorem makes investigation of the pressure-velocity relation
irrelevant to the purpose of deriving equation for the flame front position.

The results presented in this paper resolve the dilemma stated in the
Introduction in the case of stationary flames. There remains the question
of principle whether it can be resolved in the general non-stationary case.

\begin{appendix}

\section*{Consistency check for Eq.~(27)}

After a lengthy calculation in Sec.~\ref{structure}, we obtained the
following simple expression for the vorticity mode near the flame front
\begin{eqnarray}\label{a1}
v^{v}_i = \int\limits_{F} d l~\chi(\varepsilon_{pq}r_p v_{q+})
\frac{v^n_+\sigma_{+}v_{i+}}{2 v^2_+}\,.
\end{eqnarray}
\noindent
As this important formula plays the central role in our investigation,
a simple consistency check will be performed here, namely, we will
verify that $v^{v}_i$ given by Eq.~(\ref{a1}) satisfies
\begin{eqnarray}\label{check}
\left(\partial u^v/\partial\eta -
\partial w^v/\partial\xi\right)_+ \equiv \sigma_+\,.
\end{eqnarray}

\begin{figure}
\hspace{2cm}\epsfxsize=12cm\epsfbox{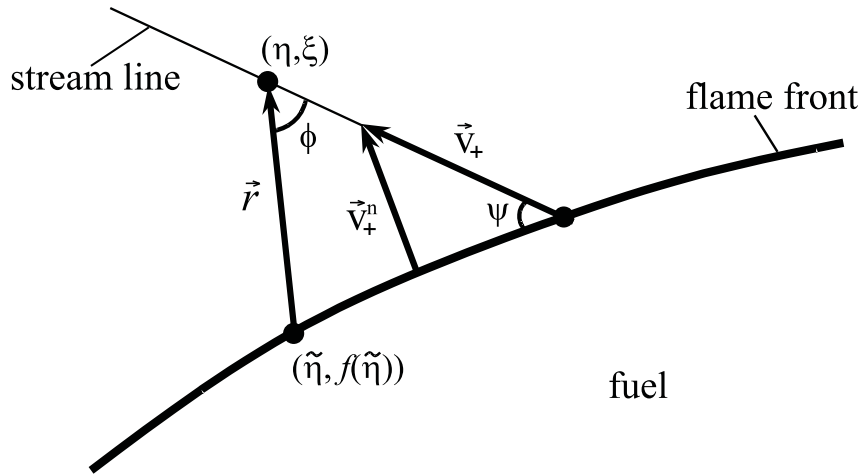}
\vspace{1cm}
\caption{Near-the-front structure of the flow downstream.
${\bf v}^n_+ = {\bf n} v^n_+$ is the normal component of the velocity.
Since the observation point $(\eta,\xi)$ is close to the flame front,
the stream line and the part of the front near this point can
be considered straight.}
\label{fig2}
\end{figure}

Contracting Eq.~(\ref{a1}) with $\varepsilon_{ki}\partial_k,$ and using
relation (\ref{deltad}),
one finds
\begin{eqnarray}\label{a2}
\varepsilon_{ki}\partial_k v^{v}_i = \int\limits_{F} d l
~\delta(\varepsilon_{pq}r_p v_{q+})\varepsilon_{ki}\varepsilon_{km} v_{m+}
\frac{v^n_+\sigma_{+}v_{i+}}{v^2_+} = \int\limits_{F} d l
~\delta(\varepsilon_{pq}r_p v_{q+})v^n_+\sigma_{+}\,.
\end{eqnarray}
\noindent
The argument of the $\delta$-function turns into zero when the vectors
$r_i$ and $v_{i+}$ are parallel. Near this point, one can write
$$\varepsilon_{pq}r_p v_{q+} \approx r v_{+}\phi\,,$$
where $\phi$ is the angle between the two vectors. On the other hand,
the line element, $d l,$ near the same point can be written as
$$d l = \frac{r}{\sin\psi} d\phi = r d\phi \frac{v_{+}}{v^n_+}\,,$$
as a simple geometric consideration shows, see Fig.~\ref{fig2}.


Substituting these expressions into Eq.~(\ref{a2}), and taking
into account relation $$\delta(\alpha x)
= \frac{1}{|\alpha|}\delta(x)\,,$$ we finally arrive at the desired
identity
\begin{eqnarray}\label{a3}
\varepsilon_{ki}\partial_k v^{v}_i
= \int\limits_{F} d\phi~\frac{r v_{+}}{v^n_+}
~\delta(r v_{+}\phi)v^n_+\sigma_{+}
=  \int\limits_{F} d\phi ~\delta(\phi)\sigma_{+} = \sigma_{+}\,.
\end{eqnarray}
\noindent
It should be noted in this respect that the identity Eq.~(\ref{check}) is
only a necessary condition imposed on the field $v^{v}_i.$ Playing the role
of a ``boundary condition'' for the vortex mode, Eq.~(\ref{check}) is
satisfied by infinitely many essentially different fields, {\it i.e.,}
fields which are not equal up to a potential. It is not difficult to verify,
for instance, that the velocity field defined by
\begin{eqnarray}
\tilde{v}^{v}_i = \int\limits_{F} d l~\chi(\varepsilon_{pq}r_p n_{q})
\frac{\sigma_+ n_{i}}{2}
\nonumber
\end{eqnarray}
\noindent
also satisfies Eq.~(\ref{check}), and the difference
$v^{v}_i - \tilde{v}^{v}_i$ is essentially non-zero.

By the construction of Sec.~\ref{structure}, $v^{v}_i$ given by
Eq.~(\ref{a1}) is essentially the only field that satisfies the
flow equations (\ref{flow1})-(\ref{flow2}).

\end{appendix}

\end{document}